\documentclass[twocolumn,aps,prr,floatfix]{revtex4-1}

\usepackage{graphicx}
\usepackage{epstopdf}
\usepackage{amsmath}
\usepackage{latexsym}
\usepackage{amsmath}
\usepackage{esdiff}
\usepackage{xcolor}

\newcommand{\beq}{\begin{equation}}
\newcommand{\eeq}{\end{equation}}

\begin{document}

\title{Quantum dynamics of single-photon detection using functionalized quantum transport electronic channels}

\author{Catalin D. Spataru}
\thanks{Corresponding author.\\ cdspata@sandia.gov}
\author{Fran\c{c}ois L\'{e}onard}

\affiliation{Sandia National Laboratories, Livermore, California 94551, USA}

\begin{abstract}
Single photon detectors have historically consisted of macroscopic-sized materials but recent experimental and theoretical progress suggests
new approaches based on nanoscale and molecular electronics. Here we present a theoretical study of photodetection in a system composed
of a quantum electronic transport channel functionalized by a photon absorber. Notably, the photon field, absorption process, transduction mechanism,
and measurement process are all treated as part of one fully-coupled quantum system, with explicit interactions. Using non-equilibrium, time-dependent 
quantum transport simulations, we reveal the unique temporal signatures of the single photon detection process, and show that the system can 
be described using optical Bloch equations, with a new non-linearity as a consequence of time-dependent detuning caused by the backaction from the transport channel via the dynamical Stark effect. We compute the photodetector
signal-to-noise ratio and demonstrate that single photon detection at high count rate is possible for realistic parameters by exploiting a novel non-equilibrium control of backaction.
\end{abstract}

\maketitle

\section{Introduction}
Single-photon detection has recently played a key role in addressing long-standing basic physics
questions such as Bell's theorem \cite{Giustina} and quantum teleportation \cite{Takesue}, as well as enabling potential new approaches 
for quantum information science \cite{Hadfield}. While existing single-photon detectors (e.g.~superconducting nanowires, avalanche photodiodes, photomultiplier tubes) 
show exquisite performance, a fundamental physics 
question is whether new types of detectors could provide even better performance. 

One avenue is to consider arrays of nanometer scale photodetectors instead of the existing
bulk detectors. In such arrays, each element simultaneously interacts with the photon field and outputs
a signal, providing advantages for performance and photon number resolution as determined from
general considerations \cite{young:2018a}. To realize such arrays it is critical to identify physical device elements that can satisfy the stringent constraints. Several years ago, quantum dot single photon detectors were first demonstrated
and have since achieved good performance \cite{Rowea}. More recently, carbon nanotubes (CNTs) functionalized 
with molecules have been explored \cite{chromophores, Guo, Hecht}, as well as photoswitched molecular electronic systems \cite{Jia}. These nano/molecular systems are promising as array elements but such systems in the presence of light-matter interactions are out of equilibrium and
thus require careful considerations of their properties. 

Previous modeling work
focused on the impact of optical absorption by the quantum transport channel itself (e.g.~a molecule \cite{Gao}, CNT \cite{Stewart}, or graphene \cite{Leonard}.)
Here we consider a different system where an electronic quantum transport channel (e.g.~a CNT) is functionalized with a photon absorber (e.g.~a molecule). We study 
single-photon detection, where the photon field, light-matter interaction, and measurement are coupled as part of one quantum system. We develop a time-dependent, 
non-equilibrium quantum transport approach for such a system, and reveal
the intricate time dynamics of single photon detection. Furthermore, we demonstrate that the
dynamics can be captured using optical Bloch equations (OBEs) with a new non-linear
contribution due to time-dependent detuning. Our approach is used to demonstrate a high signal-to-noise ratio
for single photon detection at high count rate ($\sim$GHz).

\section{Device system}

\begin{figure}
\includegraphics[trim=-20 0 -20 0,clip,width=\columnwidth]{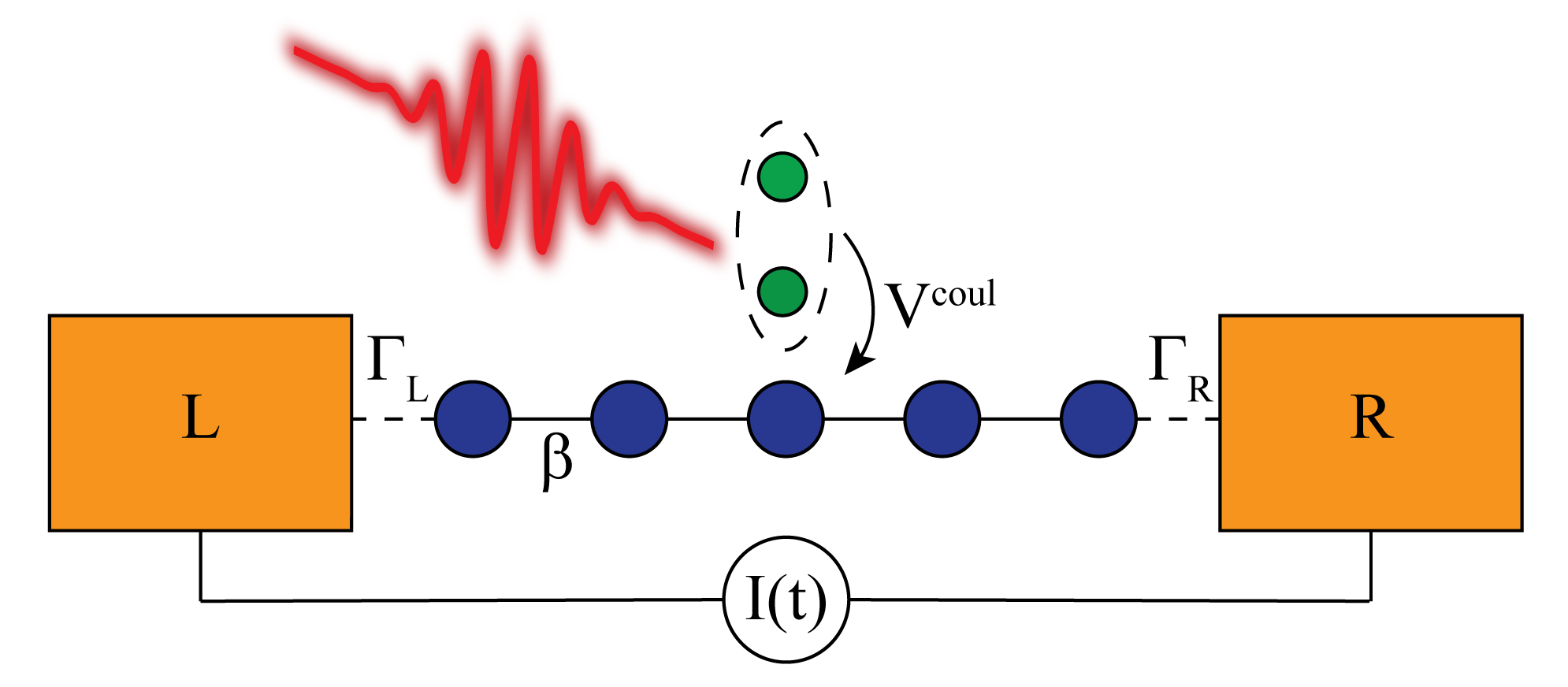}
\caption{Sketch of a device consisting of a quantum transport channel functionalized with a photon absorber. Relevant tight binding parameters are indicated with greek letters. $V^{coul}$ is the Coulomb interaction between the absorber and the transport channel.}
\vspace{-0.5cm}
\label{sketch}
\end{figure}

The system is presented schematically in Fig.~\ref{sketch}(a). 
A single-photon coherent state pulse impinges on a quantum transport channel functionalized
with a photon absorber. Upon photon absorption, the absorber acquires a permanent dipole
moment that creates an electrostatic potential in the transport channel, changing the current. 
That such a system might be able to detect single photons is not obvious because of quantum backaction.
Indeed, because the quantum transport channel interacts coherently with the molecule, measuring the
current is akin to performing a continuous quantum measurement on the molecule. Backaction from such quantum measurements have been shown to impact single photon detection in simplified models \cite{young:2018b}.

We assume that light absorption takes place in the strong-focusing regime \cite{Leuchs,Scarani,Combes} with the absorber placed in the focal point of a parabolic mirror and the spatial shape of the pulse matching the absorber dipole pattern \cite{waveguide}. 
The absorber is a diatomic system with one electron and two energy levels: the ground state (g) and the excited state (e). The ground state has no permanent dipole.
The photon energy $\hbar \omega$ is resonant with the absorber while the channel is not sensitive to the incoming light pulse. (e.g., a small diameter CNT has well-separated optical absorption peaks due to excitonic effects \cite{Spataru_CNT,Capaz_trends_Ohno}; we assume that the molecule optical gap is not matched to these peaks.) This single element detector is narrowband, but an array of such elements
could be broadband or even perform energy resolution.

The Hamiltonians $H_{abs}^0$ and $H_{ch}^0$ that describe the isolated absorber and transport channel are treated within tight-binding (TB). We assume that a single band is relevant for electronic transport. Absorber-channel coupling occurs via the Coulomb interaction, with no charge transfer between them.
The transport channel is composed of $N$ sites with inter-site distance $a=5$ bohr and hopping integral $\beta=0.5$ eV. The left (L) and right (R) leads are modeled in the wide-band limit and their coupling to the channel induces broadenings $\Gamma_{L,R}=\beta$. Parameters for the absorber were chosen
to be representative of molecules. To simulate an absorber with a realistic excited state radiative lifetime\cite{ns_tau},
we use 
$H_{abs}^0=\left[ {\begin{array}{cc}
   3  & 2.24 \\
   2.24 & -3 \\
  \end{array} } \right]$ eV and inter-site distance $d_{abs}=5$ bohr ($d_{abs}$ determines the strength of the dipole as an electron moves from one absorber site to the other). $H_{abs}^0$ includes the potential from a positive ionic charge distribution to make the absorber non-polar in the ground state. $H_{abs}^0$ results in a bare optical gap $E_g^0 = 7.5$ eV, transition dipole $d_{ge}=3.8$ Debye, and overall permanent dipole in the excited state of 10 Debye. The excited state spontaneous emission decay rate is \cite{Loudon,Steck} $\gamma_{rad}\equiv (E_g^0/\hbar c)^3d_{ge}^2/(3\pi\epsilon_0)$ which yields a radiative lifetime $\tau_{rad}\equiv\hbar/\gamma_{rad}=1$ ns. This radiative lifetime is the natural reset that determines the count rate of the detector $\sim$GHz.
  
The Coulomb potential between absorber and channel is described via the Ohno parameterization \cite{Ohno,Capaz_trends_Ohno} as  
$V^{Coul}_{ij}=U/\sqrt{1+\left[(4\pi\epsilon_0/e^2)Ur_{ij}\right]^2}$,
where $r_{ij}$ is the absorber-channel inter-site distance and $U=5$ eV. The separation between the channel and the farthest absorber site is $d=20$ bohr, and the permanent dipole is oriented perpendicular to the channel to maximize the impact on electron transport.
The Coulomb coupling may lead to near-field non-radiative energy transfer between the absorber and the transport channel, which could prevent the absorber from getting excited. 
We minimize this effect by having the absorber optical gap larger than the channel band width $4\beta$, such that no intra-channel excitations match the absorber de-excitation energy \cite{substrate_loss}. 

We model electrostatic doping of the channel \cite{Spataru_dop} via a fictitious gate potential $V_G$ that acts on the on-site energies of the channel. 
This is equivalent to a potential of opposite sign acting on the leads while leaving the channel unaffected, so the Fermi level inside the two leads depends on $V_G$ as well as the applied bias voltage $V_{sd}$: $E_F^{L,R}=\pm V_{sd}/2 -V_G$.
Doping the channel creates a net charge, producing an electric field that slightly polarizes the ground state of the absorber and renormalizes the optical gap of the absorber from $E_g^0$ to $E_g$. The photon energy is $\hbar\omega =E_g+\Delta$ with $\Delta$ a possible small detuning.

\section{Results}
\subsection{NEGF dynamics}
To study the system dynamics, we employ a non-equlibrium Green's function (NEGF) formalism based on the equation of motion for the 1-particle Green's function \cite{Stefanucci}, treating the Coulomb interaction inside the functionalized transport channel at the mean-field Hartree-Fock level (see Appendix A for formalism details). 
We also performed extensive calculations where electron correlation effects were included within the $GW$ approximation \cite{Louie}, as discussed in Appendix B.

The equation of motion for the density matrix of the coupled absorber-channel system $\rho$ is projected (i) on the absorber to give $\rho_{abs}= 
\left[ {\begin{array}{cc}
   \rho_{gg} & \rho_{ge} \\
   \rho_{eg} & \rho_{ee} \\
  \end{array} } \right]$
in the eigenstate basis of $H_{abs}^0$ subject to the initial electric field generated by the doped channel, 
and (ii) on the channel   to give $\rho_{ch}$, a $N \times N$ matrix. Their {coupled} dynamical equations are:
\begin{multline}
 \frac{\partial\rho_{abs}}{\partial t}+i[H_{abs},\rho_{abs}]=\gamma^{rad}
\left[ {\begin{array}{cc}
   \rho_{ee} & -\rho_{ge}/2 \\
   -\rho_{eg}/2 & -\rho_{ee} \\
  \end{array} } \right]
\label{coupled_rho_MOL}
\end{multline}
\begin{multline}
 \frac{\partial\rho_{ch}}{\partial t}+i\{H_{ch}\rho_{ch} -{\it{h.c.}}\}= \\
 \int{d\tilde{t}} \{G_{ch}(t,\tilde{t})\Sigma^<_{L+R}(\tilde{t}-t)+{\it{hc}}\}
\label{coupled_rho_CNT}.
\end{multline}
The channel Green's function is given by
\beq
G_{ch}(t,\tilde{t})=i\theta(\tilde{t}-t) exp\left[-i\int_{\tilde{t}}^t{dt'} H_{ch}(t') \right]
\label{ret_G_ch}
\eeq
while $\Sigma^<_{L,R}(t)=i\Gamma_{L,R}\int{dE} e^{-itE}f_{L,R}(E)$ are the lead self-energies, and $f_{L,R}$ is the Fermi-Dirac electron distribution in the left/right lead. The effective Hamiltonians are
\beq
H_{abs}=H^0_{abs}+\Sigma_{abs}^{Ha}(\rho)+{\bf{r}}\cdot{\bf{E}}(t)
\label{H_abs}
\eeq
\beq
H_{ch}=H^0_{ch}+\Sigma_{ch}^{Ha}(\rho)-i\Gamma_{L+R}/2
\label{H_ch}
\eeq
where the Hartree self-energy depends self-consistently on $\rho$:
$\Sigma^{Ha}_{ii}(t)=\sum_j^{abs+ch} \rho_{jj}(t) V_{ji}^{Coul}$.
The last term in Eq.~\eqref{H_abs} is the coherent field term which depends on the electric field ${\bf{E}}$ associated with the single-photon. ${\bf{E}}$ is polarized parallel to the molecule while its temporal shape is determined by the pulse envelope $\xi$: $E(t)=2\sqrt{\gamma_{rad}}/d_{ge} \times e^{i\omega t}\xi(t)$. We choose a Gaussian pulse $\xi(t)^2=\Omega_0/\sqrt{2\pi} \times \textrm{exp}[-\Omega_0^2(t-t_c)^2/2]$ with bandwidth $\Omega_0$ and centered at $t_c$. The normalization gives one photon on average in the light pulse.


\begin{figure}
\vspace{-0.5cm}
\includegraphics[trim=0 40 240 10,clip,width=\columnwidth]{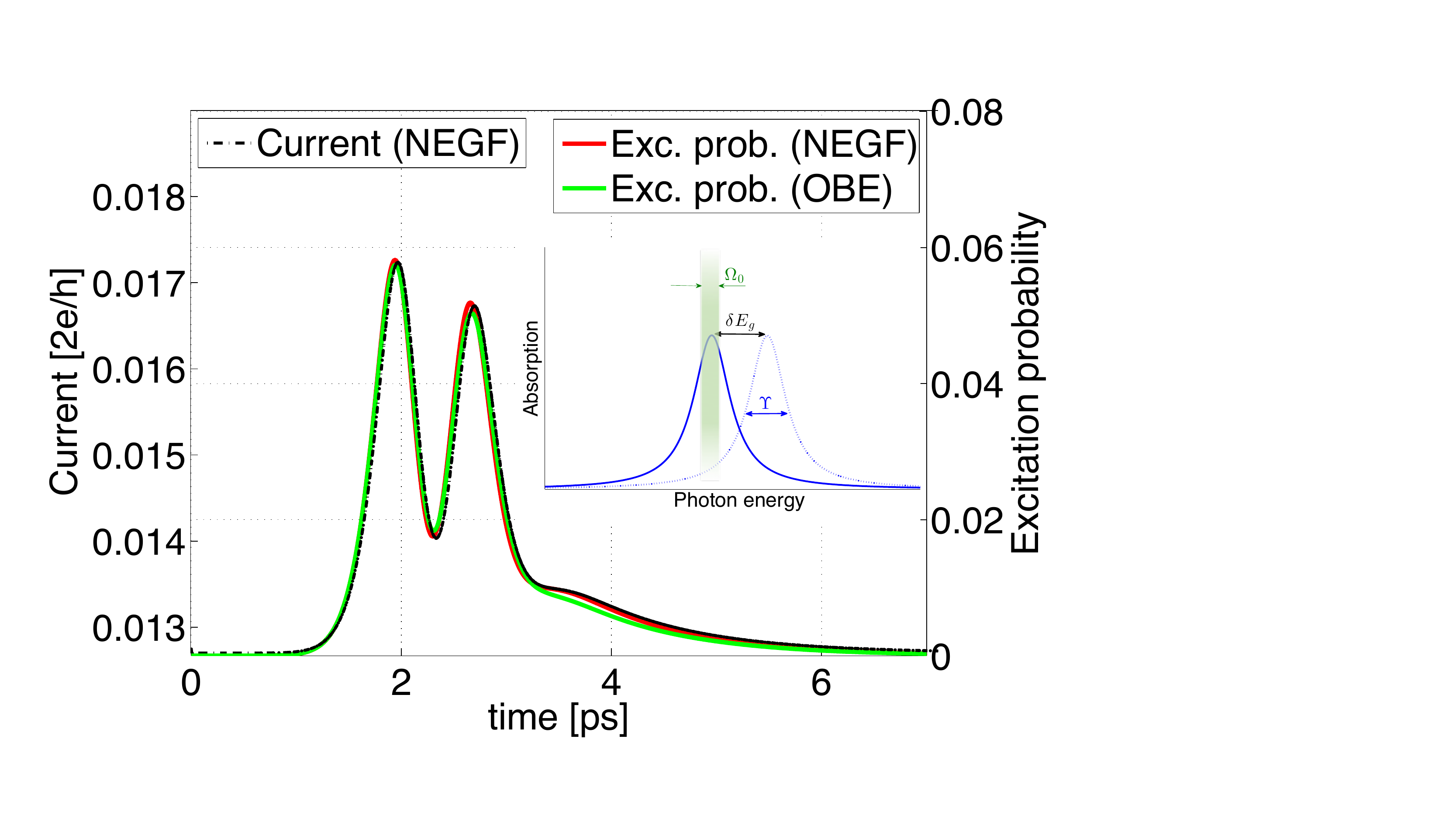}
\caption{Calculated time-dependent expectation value of the current through the device (dashed line) and the probability of finding the absorber in the excited state (solid lines). $N=3$, $\tau^{rad}=1$ ps, $\tau_0=417$ fs, {$t_c=2.33$ ps}. The inset illustrates the absorber level detuning via the dynamical Stark effect.}
\vspace{-0.0cm}
\label{example_prob_exc_current}
\end{figure}

Because of the computational cost of the NEGF formalism, we first study the case of an absorber with $\tau^{rad}$ set artificially to $1$ ps, placed above the middle of a $3$-site channel ($N=3$) subject to $V_G=1.125$ eV and $V_{sd}=0.5$ eV. We choose a photon pulse duration $\tau_0=\tau_{rad}/2.4$.
Fig. \ref{example_prob_exc_current} shows the time-dependent expectation value of the current through the device $I(t)$. 
We use the following expression for the time-dependent expectation value of the current from the left/right contacts \cite{Wingreen}:
\begin{multline}
J_{L/R}(t)=-\frac{e}{\hbar}Tr\{\rho_{ch}(t) \Gamma_{L/R}\}+\\
\frac{e}{\hbar}\int_{\bar{t}} Tr\{ 
\Sigma^<_{L/R}(\bar{t}-t)G_{ch}^r(t,\bar{t}) +{\it h.c.}.
\}
\label{JL}
\end{multline}
(We note that $J_L=-J_R$ only in the steady-state.)
We present results for the symmetrized current: $I(t) \equiv (J_L(t)-J_R(t))/2$, using a smoothing procedure (moving average filter) over a time span equal to a few light cycles. We see in Fig. \ref{example_prob_exc_current} that the current shows rich temporal behavior, with a clear increase signaling photon detection, but modulated by strong oscillations.
Fig. \ref{example_prob_exc_current} also shows the absorber excitation probability $\rho_{ee}(t)$, which correlates with the time-dependent current change $\delta I(t)=I(t)-I_0$ via $\delta I(t)=c{\times}\rho_{ee}(t)$,
with a proportionality factor $c$ that can be found via simple steady-state calculations within linear response.
The simple relationship between $\rho_{ee}(t)$  and $\delta I(t)$ is due to the fast relaxation times (order of fs) for electrons inside the channel as expected from the magnitude (order of eV) of the parameters $\Gamma_{L,R}$ and $\beta$.

The time dependence of the excitation probability can be understood in a simplified picture {(see Fig.~\ref{example_prob_exc_current} inset)} where the absorption spectrum for the absorber is dynamically modulated over time scales ($\sim$ps) much larger than the light period ($\sim$fs). In general a steady-state light field causes power broadening \cite{Loudon,Steck} giving $\Upsilon=\sqrt{\gamma_{rad}^2+2\Omega_{Rabi}^2}$ where $\Upsilon$ is the width of the molecular level absorption peak and $\Omega_{Rabi}= Ed_{ge}$ is the Rabi frequency. For a single photon pulse the upper limit for $\Upsilon$ is $\sqrt{\gamma_{rad}^2+8/\sqrt{2\pi}\gamma_{rad}\Omega_0}$. In the isolated absorber case the optimal pulse bandwidth is \cite{Scarani} $\Omega_0=2.4\times\gamma_{rad}$ which results in a similar value for $\Upsilon$ and maximizes the overlap between the power broadened absorption line and the bandwidth of the pulse leading to a maximum excitation probability of $\approx 50\%$.
In the coupled case however, the same choice for pulse duration results in smaller excitation probability, as seen in Fig. \ref{example_prob_exc_current}. This is due to the dynamical Stark effect: as the photon excites the absorber, the associated permanent dipole changes the charge distribution in the transport channel, which in turn changes the electric field acting back on the absorber. The induced time-dependent renormalization $\tilde{E}_g(t)=E_g+\delta E_g(t)$ of the optical gap leads to strong detuning $\hbar\omega-\tilde{E}_g(t)=\Delta-\delta E_g(t)$ which reduces the absorber excitation. This is illustrated in the inset of Fig.~\ref{example_prob_exc_current} where it is also {apparent} that a small detuning $\Delta$ matching the mean value of $\delta E_g(t)$ may maximize the light-absorber interaction. The time-dependent detuning also explains the oscillating behavior: as the excitation probability and the detuning increase the dynamics becomes dominated by spontaneous emission which decreases the excitation probability. Thus the dynamics is governed by a complex interplay of interactions between the absorber and the quantum transport channel.

\begin{figure}
\includegraphics[trim=-10 0 -10 0,clip,width=\columnwidth]{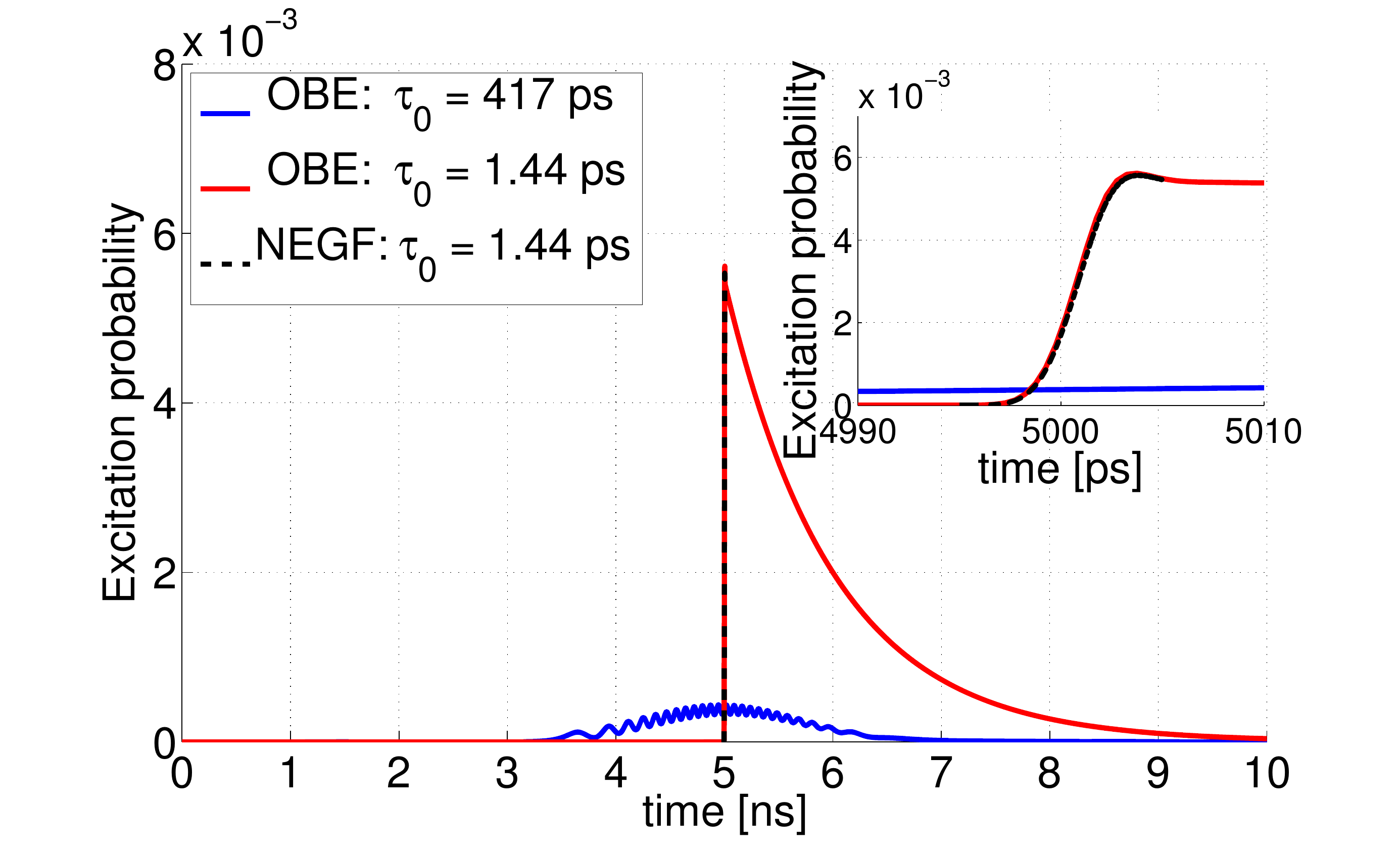}
\caption{Absorber excitation probability calculated within OBE for two different pulse durations $\tau_0$. NEGF results for the short $\tau_0$ are visible in the inset where the propagation time is zoomed-in over  few ps. $N=3$, $\tau^{rad}=1$ ns, {$t_c=5$ ns}.}
\vspace{-0.5cm}
\label{opt_pulse_dur}
\end{figure}

\subsection{Nonlinear optical Bloch equations}
Having understood the time dependence of $\rho_{ee}(t)$ we now show that we can replace the computationally demanding NEGF approach with a simpler (and computationally faster) description, allowing us to simulate longer time scales. In the case of an isolated absorber the evolution of the absorber density matrix can be written as optical Bloch equations (OBE) \cite{Steck,other_approaches}:
\begin{multline}
\diffp{\rho_{ee}}{t}=i\sqrt{\gamma_{rad}}\xi(t)\left(\tilde{\rho}_{eg}-\tilde{\rho}_{ge}\right)-\gamma_{rad} {\rho}_{ee}
\\
\diffp{\tilde{\rho}_{ge}}{t}=-i \sqrt{\gamma_{rad}}\xi(t) \left({\rho}_{ee}-{\rho}_{gg}\right)-\left(\frac{\gamma_{rad}}{2}+i\Delta \right)\tilde{\rho}_{ge}
\label{OBE}
\end{multline}
with $\rho_{gg}=1-\rho_{ee} , \ \tilde{\rho}_{eg}=\tilde{\rho}_{ge}^* $ and where we made use of the random wave transformation $\tilde{\rho}_{eg(ge)}=\rho_{eg(ge)} e^{+(-)i\omega t}$. We find that the coupled case can be described by the same set of equations but with {$\Delta$ replaced by} a time-dependent detuning $\Delta-\delta E_g(t)$ that depends on the excitation probability through $\delta E_g(t)= {f}(\rho_{ee}(t))$. The function $f$ satisfies $f(0)=0$ and can be easily fitted via a few steady-state calculations. We note that the resulting modified OBE equations are non-linear due to the term $\delta E_g \tilde{\rho}_{ge}$.
Fig. \ref{example_prob_exc_current} demonstrates the excellent agreement between the NEGF and OBE approaches. The OBE simulations (employing $\sim1$ ps timestep) in combinations with steady-state calculations of $c$ and ${f}$ yield a computational time $\sim5$ orders of magnitude faster than their NEGF counterpart (which uses $\sim10$ attosecond time-step), being thus well suited to search in the multi-dimensional parameter space for efficient single-photon detection. 

One important parameter is the duration of the light pulse (bandwidth). 
Indeed, the light-matter coupling is not optimal when the photon bandwidth is much smaller than the optical gap renormalization $\delta E_g$. The optimized coupling is realized when the renormalized optical level remains within the bounds set by the photon bandwidth during the course of the light-absorber interaction. This is demonstrated in Fig.~\ref{opt_pulse_dur} which shows OBE results for the excitation probability of the absorber with its actual $\tau_{rad}=1$ ns coupled to the $3$-site channel described above, for two different pulse durations: i) $\tau_0=\tau_{rad}/2.4=417$ ps which is optimal for an isolated absorber; ii) $\tau_0=1.44$ ps which is optimal in the coupled case. Case ii) yields a larger $\Omega_0$ and $\Upsilon$ which decreases the impact of detuning (see Fig.~2 inset). One can see that the excitation probability is more than an order of magnitude larger in the latter case. Also shown is the result of a full NEGF calculation, which could only be run for a few ps, but which shows excellent agreement with the OBE calculation. We note that after the pulse passes, the molecule relaxes by spontaneous emission, leading to a short detector reset time $\sim 1$ ns. We find that the parameter $\Delta$ is less impactful since the optimal value increases the excitation probability by $<30\%$ w.r.t the $\Delta=0$ case. 

\begin{figure}
\includegraphics[trim=0 30 0 0,clip,width=\columnwidth]{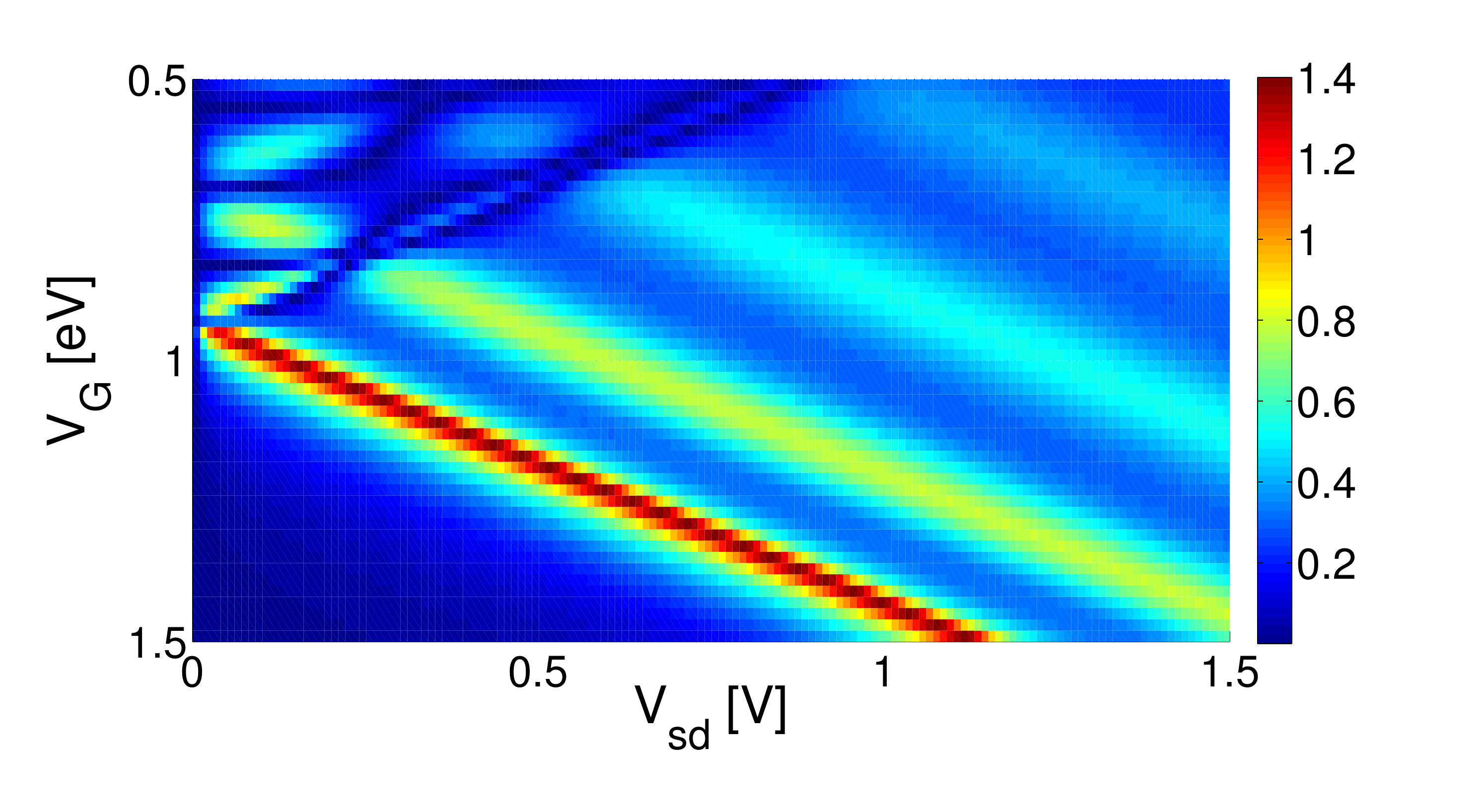}
\caption{SNR for an absorber placed above the middle of a $7$-site channel. $\tau_0$ and $\Delta$ are optimized for each ($V_{sd}$, $V_G$) set. The time window is $1.2$ ns and starts when the photon pulse begins interacting with the absorber. $\tau^{rad}=1$ ns.}
\vspace{-0.5cm}
\label{SNR_7sites}
\end{figure}

\subsection{Signal-to-noise ratio}

During the interaction between the photon pulse and the coupled absorber-channel system the expectation value of the current $I(t)$ changes dynamically from its steady-state value $I_0$ prior to light interaction. The main question we ask is: can {the change} $\delta I(t)$ be detected? Answering this question requires a calculation of the signal-to-noise ratio (SNR).
In this work we consider the situation where the temperature is $T=0$ and where the detector can be operated at high frequency (this is justified {below}); this implies that the thermal noise and $1/f$ noise are negligible, so the only source of noise is the shot noise. Assuming a Poisson distribution of the passing electrons, the root mean square of the current fluctuations in the steady state is $\sigma_{I_0} = \sqrt{n_0}$ where $n_0$ is the integral of $I_0$ over a certain time window, i.e.~$n_0$ is the number of charge carriers that flow in the time window. We also assume a conservative value for the Fano factor, namely $F=1$ \cite{Fano}. We then define SNR $= {\delta n}/{\sqrt{n_0}}$ where $\delta n$ is the integral of $\delta I$ over the time window.

We find that for a given geometry the optimized SNR is achieved for $V_{sd}$ and $V_G$ such that one Fermi level ({\it e.g.} $E_F^R$) pins the highest prominent peak in the density of states (DOS) of the channel while the other Fermi level is much higher ($E_F^L\gg E_F^R$). 
Fig. \ref{SNR_7sites} shows SNR results for the absorber placed above ($d=20$ bohr) the middle ($j=4$) of a $N=7$ channel for a range of bias and gate voltages. The peak values in the SNR correlate with resonances in the channel DOS. The overall SNR decreases as $E_F^R$ moves towards the lower energy DOS peaks due to the increasing number of transport eigenchannels inside the bias window that are not impacted by the absorber-channel interaction. For this geometry we obtain an optimized SNR of 1.4 (with $E_F^R=0.92$ eV, $\tau_0=372$ fs and an optimal time window of $1.2$ ns starting when the photon pulse begins interacting with the absorber).  

\begin{figure}
\includegraphics[trim=0 0 0 0,clip,width=\columnwidth]{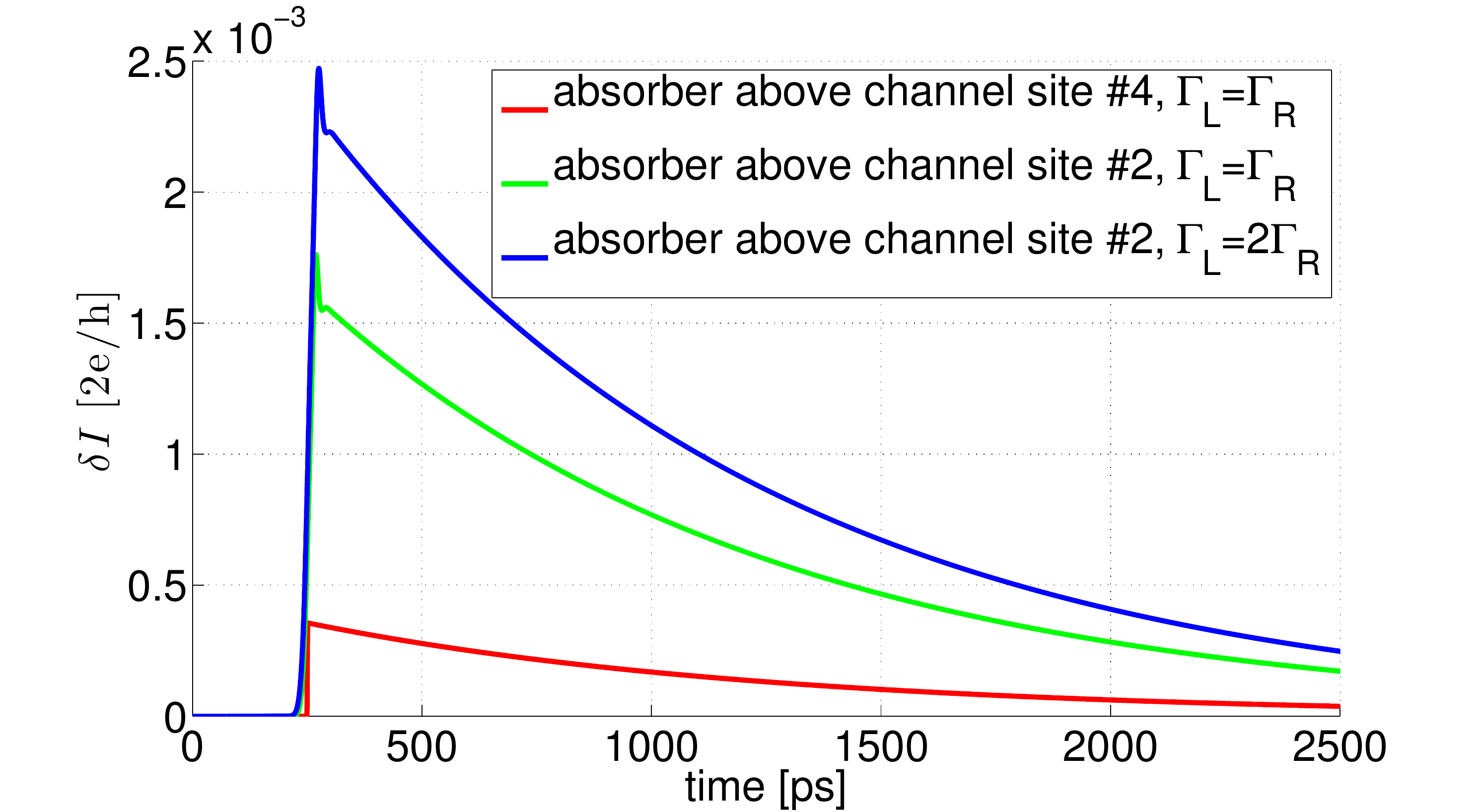}
\caption{OBE results for the change in current $\delta I(t)$ for three different design strategies. $N=7$, $E_F^L\gg E_F^R$, $\tau^{rad}=1$ ns, {$t_c=250$ ps}. The value  of the current prior to light interaction is $I_0[2e/h]=$0.013, 0.043 and  0.054 respectively.}
\vspace{-0.5cm}
\label{current_3designs}
\end{figure}

\begin{figure}
\includegraphics[trim=40 100 40 100,clip,width=\columnwidth]{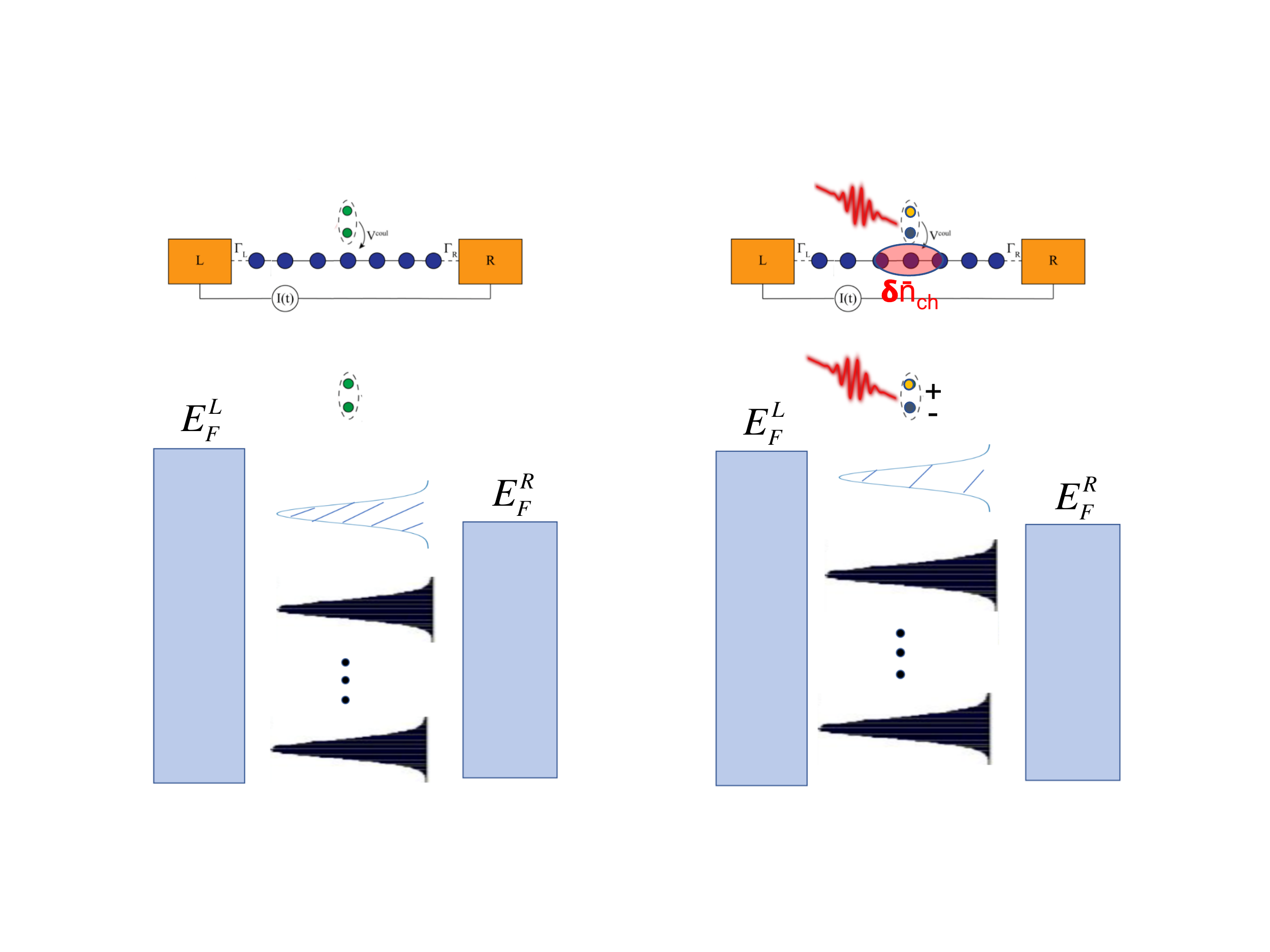}
\caption{Upper panels: sketch of the functionalized channel device before (left) and during (right) photoexcitation. Lower panels: energy level alignment between channel and leads for a possible bias and gate voltage configuration where only the highest energy channel level contributes to the channel current/is positioned near the bias window. Partial level occupancy is indicated with hashing. As the channel energy levels shift up upon absorber photoexcitation (right panel), the electron occupancy of the highest energy level decreases, as indicates with less hashing. The resulting depopulation $\delta \bar{n}_{ch}$ in the channel electronic density modifies the electric field induced by the channel and impacting the absorber.}
\vspace{-0.5cm}
\label{action_backaction}
\end{figure}

For efficient photodetection one would like to achieve larger SNR values. Higher SNR is possible through non-equilibrium control of backaction.  Indeed, Fig.~\ref{current_3designs} shows that when the molecule is placed closer to the left electrode ($j=2$), the change in current is significantly higher and the optimized SNR increases to 3.5 (with $E_F^R=0.88$ eV and $\tau_0=10$ ps).
The reason for the improved SNR in the asymmetric case is a reduction of the channel backaction on the molecule by a subtle non-equilibrium electronic transport effect. Indeed, the electronic transport is dominated by a narrow energy level (index $N$) located near $E_F^R$. When the molecule is closer to the left electrode, the asymmetric Coulomb perturbation created by the excited molecule increases the coupling  between the left lead and the channel level to $\tilde{\Gamma}_L^N=\Gamma_L^N(1+\delta)$ and decreases its coupling to the right lead to $\tilde{\Gamma}_R^N\approx\Gamma_R^N(1-\delta)$. This results in an increase {$\delta \tilde{n}_{ch}$} of the non-equilibrium occupancy \cite{Wingreen} of the level which for symmetric leads ($\Gamma_L^N=\Gamma_R^N$) is 
$\delta \tilde{n}_{ch} \approx \delta/2 \int dE [f_L(E)- f_R(E)] \textstyle{DOS}(E)$, where the factor $f_L- f_R$ indicates that $\delta \tilde{n}_{ch}{\ne0}$ is essentially a non-equilibrium effect.
This effect opposes the more general decrease in level occupancy {$\delta \bar{n}_{ch}$} generated when the electron in the absorber gets closer to the channel pushing away its electrons into the leads (as illustrated in Fig. \ref{action_backaction}).
Thus, the electric field causing detuning of the absorber is smaller in the asymmetric case, 
ultimately allowing much higher excitation probability ($3.3\%$ for $j=2$ versus $0.17\%$ for $j=4$). 

Further increase of SNR is possible in configurations where the leads are asymmetrically coupled to the channel before photoexcitation. Indeed, as seen in Fig.~\ref{current_3designs}, {a larger change in current} is achieved by coupling the level more strongly to the left lead by making $\Gamma_L = 2\Gamma_R=1 $eV. For the optimal parameters
$E_F^R=0.86$ eV and $\tau_0=12$ ps the absorber excitation probability is $4.0\%$ and during the $1.2$ ns time window there are about $800$ more electrons passing through the device{, leading to an SNR of 4.5}.

\section{Conclusion}

In conclusion, we performed quantum-mechanical calculations of single-photon detection for an absorber coupled to a quantum transport channel via the Coulomb interaction. Notably, our work considers
the coherent photon field, optical absorption process, and transduction mechanism as part of one coupled quantum system. 
We reveal the complex dynamics of such systems, and identify non-equilibrium electronic transport as a mechanism to control backaction. With a purely electronic coupling between the absorber and transport channel we find a signal-to-noise ratio larger than $4$ and with reset time of $\sim 1$ ns which suggests that single-photon detection is possible with a minimal device design.  This SNR arises for optimal pulse durations of $\sim 10$ ps; it is possible to increase the optimal pulse duration by decreasing the absorber-channel coupling ({\it e.g.} increasing the separation), or by 
using absorbers with longer spontaneous emission decays at the cost of lower count rate. We anticipate rich additional physics in these systems, for example through considerations of current fluctuations and full counting statistics of electrons \cite{noise,FCS,Nazarov}. More broadly, we envision these systems as part of large arrays of nanoscale detection elements, all simultaneously interacting with the photon field.
\renewcommand{\theequation}{A-\arabic{equation}} 
\setcounter{equation}{0}

\section*{Appendix A: Quantum kinetic equation for the one-particle density matrix}

Here we derive the quantum kinetic equation for the one-particle charge density matrix $\rho$ of the functionalized transport channel interacting with a single photon coherent light field.
We use a non-equilibrium Green's function (NEGF) approach for the one-particle Green's function:
\beq
G(1,2)=-i\langle T_C \left[\Psi_H(1)\Psi^\dagger_H(2)\right]\rangle
\eeq 
where $G$ is the Green's function, $\Psi$ is the system wavefunction for Hamiltonian $H$, and $T_C$ denotes time ordering along the double-time Keldysh contour \cite{Keldysh}. The arguments for $G$ denote temporal, spatial and spin degrees of freedom. For simplicity, next we show explicitly only the time arguments. 
One starts from the Dyson equation:
\beq
(G_0^{-1}-\Sigma)G=I
\label{Dyson}
\eeq
where the self-energy $\Sigma$ accounts for electron-electron (Coulomb) interactions inside the functionalized transport channel as well as for the coupling of the transport channel to the leads. In this equation, $G_0$ is the Green's function for the isolated functionalized channel in the absence of leads. $G_0$ is diagonal in the time-domain:
\beq
G_0^{-1}(t_1,t_2)=\left[ i{\partial t_1} -H^0_{tot} -U(t_1)\right]\delta(t_1-t_2)
\label{G0}
\eeq
with $H^0_{tot}=H^0_{abs}+H^0_{ch}$ the single-particle time-independent Hamiltonian of the functionalized channel 
(bare, with no electron-electron interation). $U(t)$ is the radiation field which at this point is treated classically within the dipole approximation \cite{Attaccalite}: $U(t)=-\vec{d} \cdot \vec{E}(t)$ where $\vec{d}\equiv e\vec{r}$ is the dipole operator. 

The Dyson equation leads to the Kadanoff-Baym equations (KBEs) -a set of coupled 
integro-differential equations for the lesser/greater/advanced/retarded Green's functions involving two-time arguments $G^{</>/a/r}(t_1,t_2)$. Using the Langreth rules \cite{Langreth} for analytic continuation, the KBEs read \cite{Haug_book,Bonitz}:
\begin{multline}
\left[ i{\partial t_1} -H^0_{tot} -U(t_1)\right]G^r(t_1,t_2)=\\
\delta(t_1-t_2)+\int dt \ \Sigma^r(t_1,t)G^r(t,t_2) 
\label{KBE_Gr}
\end{multline}
\begin{multline}
\left[ i{\partial t_1} -H^0_{tot} -U(t_1)\right]G^<(t_1,t_2)=\\
\int d\bar{t} \ \{\Sigma^r(t_1,\bar{t})G^<(\bar{t},t_2)+\Sigma^<(t_1,\bar{t})G^a(\bar{t},t_2)\} 
\label{KBE_Gless1}
\end{multline}
\begin{multline}
\left[ -i{\partial t_2} -H^0_{tot} -U(t_2)\right]G^<(t_1,t_2)=\\
\int d\bar{t} \ \{ G^r(t_1,\bar{t})\Sigma^<(\bar{t},t_2)+G^<(t_1,\bar{t})\Sigma^a(\bar{t},t_2)\} 
\label{KBE_Gless2}
\end{multline}
and similar equations for the other Green's function components \cite{Haug_book,Bonitz}.

Subtracting Eqs.~\ref{KBE_Gless1} and \ref{KBE_Gless2}  and setting $t_1=t_2=t$ one obtains the Generalized Kadanoff-Baym (GKB) equation \cite{Spicka,GKBref} for $\rho(t)\equiv \langle \Psi^\dagger_H(t) \Psi_H(t) \rangle =-i G^<(t,t)$:
\begin{multline}
\frac{\partial \rho}{\partial t}+i\left[H^0_{tot}+U(t),\rho(t)\right]_-=
\int_{t_0}^t d\bar{t} \
\{
\Sigma^<(t,\bar{t})G^a(\bar{t},t)-\\
G^r(t,\bar{t})\Sigma^<(\bar{t},t)
+\Sigma^r(t,\bar{t})G^<(\bar{t},t)-G^<(t,\bar{t})\Sigma^a(\bar{t},t)
\}
\label{GKB_rho}
\end
{multline}
with the initial condition formulated at the reference time $t_0$.

So far the GKB equation does not represent a closed equation for $\rho$ because the r.h.s. of Eq. \ref{GKB_rho} involves Green's functions with two different time arguments. Solving the KBE for $G^{</r}$ via a two-time propagation approach is a formidable task which has been achieved only in very simple cases within certain approximations for $\Sigma$. Most studies rely on an approximation ({\it e.g.} GKB Ansatz (GKBA) \cite{Lipavski}) to reconstruct two-time Green's functions from $\rho$ which results in closure of the GKB equation. In this work we do not need to invoke the GKBA as we achieve single-time propagation of $\rho$ using the following simplifications.
First, we treat Coulomb interaction effects via  first-order perturbation in the Coulomb potential, {\it i.e.}~at the mean-field Hartree-Fock level \cite{HF}. Second, we consider that the leads are well described by the wide-band limit. 
These two approximations simplify the self-energy expressions:
\beq
\Sigma^{r/a}(t,\bar{t})=\left[\Sigma^{Ha}(t)\mp i\left(\Gamma_L+\Gamma_R\right)/2\right] \delta(t,\bar{t})
\label{SigmaR}
\eeq
\beq
\Sigma^<(\bar{t},t)=\Sigma_L^<(\bar{t}-t)+\Sigma_R^<(\bar{t}-t)
\label{SigmaLess}
\eeq
where $\Sigma_{L/R}^<$ are simply related to the Fourier transform of the Fermi-Dirac distribution functions in the left ($L$) and right ($R$) leads via the energy-independent broadenings $\Gamma^{L/R}$ induced by the contacts \cite{Wingreen}:
\beq
\Sigma_{L/R}^<(\tau)= 
i\Gamma^{L/R}\int \frac{dE}{2\pi} e^{-iE\tau} f_{L/R}(E).
\label{SigmaLess2}
\eeq
The Hartree self-energy depends self-consistently on the site-diagonal part of $\rho$ and reads (in the site basis):
\beq
\Sigma^{Ha}_{ii}(t)=\sum_j \rho_{jj}(t) V_{ji}^{Coul}.
\label{SigmaHF}
\eeq

The aforementioned simplifications result in closure of the GKB equation \cite{Stefanucci}:
\begin{multline}
\frac{\partial \rho}{\partial t}+i\{\left[H^0_{tot}+\Sigma^{Ha}(t)+U(t)-i\left(\Gamma_L+\Gamma_R\right)/2\right]\rho(t) -{\it h.c.}\}=\\
-\int_{t_0}^t d\bar{t} 
\{
G^r(t,\bar{t})
\left[ \Sigma_L^<(\bar{t}-t)+\Sigma_R^<(\bar{t}-t) \right]
+{\it h.c.}
\}.
\label{GKB_close}
\end{multline}
where $G^{r}$ obeys the following simplified version of the nonequilibrium Dyson Eq.~\ref{KBE_Gr}:
\begin{multline}
\left[ i{\partial t_1} -H^0_{tot} -\Sigma^{Ha}(t_1)-U(t_1)-
i\left(\Gamma_L+\Gamma_R\right)/2\right]\times \\
G^r(t_1,t_2)=\delta(t_1-t_2) 
\label{KBE_Gr_simple}
\end{multline}
with solution
\begin{multline}
G^{r}(t,\bar{t})=-i\theta(t-\bar{t})\times \\
exp\{-i\int_{\bar{t}}^t dt'
\left[H^0_{tot}+U(t')+\Sigma^{Ha}(t')- i\left(\Gamma_L+\Gamma_R\right)/2\right]\}.
\label{GR}
\end{multline}

It is convenient to take advantage of the absence of charge transfer between absorber and transport channel. For that, we first note that if no initial correlations exist between the two subspaces at the initial time $t_0$ then one has at any subsequent time $t$:
\beq
\rho(t)=
\left[ {\begin{array}{cc}
   \rho_{abs}(t)  & 0 \\
   0 & \rho_{ch}(t) \\
  \end{array} } \right] 
\eeq
We also note that the broadenings $\Gamma^{L/R}$ exist only in the channel subspace. Projecting Eqs \ref{GKB_close} and \ref{GR} on the absorber and transport channel subspaces we obtain:
\beq
\frac{\partial \rho_{abs}}{\partial t}+i[H^0_{abs}+\Sigma_{abs}^{Ha}(t)+U(t),\rho_{abs}(t)]_-=0,
\label{rho_MOL}
\eeq
\begin{multline}
\frac{\partial \rho_{ch}}{\partial t}+
i\{\left[H^0_{ch}+\Sigma_{ch}^{Ha}(t)-i\left(\Gamma_L+\Gamma_R\right)/2\right]\rho(t) -{\it h.c.} \}=\\
-\int_{t_0}^t d\bar{t} 
\{
G^r_{ch}(t,\bar{t})
\left[ \Sigma_L^<(\bar{t}-t)+\Sigma_R^<(\bar{t}-t) \right]
+{\it h.c.},
\}
\label{rho_CNT}
\end{multline}
\begin{multline}
G_{ch}^{r}(t,\bar{t})=-i\theta(t-\bar{t})\times \\
exp\{-i\int_{\bar{t}}^t dt' \left[H_{ch}^0+\Sigma_{ch}^{Ha}(t)- i\left(\Gamma_L+\Gamma_R\right)/2\right]\}.
\label{GR_CNT}
\end{multline}

Equation \ref{rho_MOL} shows that the electron dynamics in the absorber differs from the isolated absorber case only through the time-dependent potential $\Sigma_{abs}^{Ha}(t)$ generated by the charge in the channel in response to self-consistent changes in the absorber charge density upon photoexcitation. The absorber dynamics is thus effectively equivalent to that of an isolated two-level system subject to a modulated (no longer mono-chromatic) light field.

So far we have treated the radiation field classically. In the case of coherent light (as opposed to a Fock state)  this classical treatment -together with a proper normalization of the electric field entering the expression for $U(t)$- accounts for the absorbtion and stimulated emission components \cite{Scarani,Combes} of the light-absorber interaction. However, the spontaneous emission component-which cannot be treated classically- is crucial and needs to be accounted for in the context of single photon detection. In order to introduce it, we note that 
this component should describe the decay of the absorber excited state to the ground state via spontaneous emission of radiation into the environment (vacuum) independently of the occupation of the quantized states of the electromagnetic environment.
It is thus sufficient to consider the case of an isolated absorber in the absence of a light pulse. In this case the complete dynamics can be obtained via the master equation approach \cite{Combes} or from optical Bloch equations \cite{Steck} and it yields the following kinetic equation:
\beq
\frac{\partial \rho_{abs}}{\partial t}+i[H^0_{abs},\rho_{abs}(t)]_-=
\gamma^{rad}
\left[ {\begin{array}{cc}
   \rho_{ee} & -\rho_{ge}/2 \\
   -\rho_{eg}/2 & -\rho_{ee} \\
  \end{array} } \right]
\label{rho0_MOL_gamma_rad}
\eeq
where $\gamma^{rad}$ denotes the radiative decay rate of the absorber excited state and we used the following notation for $\rho_{abs}$ in the absorber ground (g) and excited (e) state basis: 
$\rho_{abs}\equiv \left[ {\begin{array}{cc}
   \rho_{gg} & \rho_{ge} \\
   \rho_{eg} & \rho_{ee} \\
  \end{array} } \right]. 
$

By comparing Eqs \ref{rho_MOL} and \ref{rho0_MOL_gamma_rad} it follows that the {\it r.h.s.}~of 
Eq.~\ref{rho0_MOL_gamma_rad} represents the spontaneous emission term. We can now include this term in the more general case that describes the absorber interacting with the transport channel and the coherent light pulse  (Eq. \ref{rho_MOL}), with the final expression reading:
\beq
\frac{\partial \rho_{abs}}{\partial t}+i[H^0_{abs}+\Sigma_{abs}^{Ha}(t)+U(t),\rho_{abs}(t)]_-=
\gamma^{rad}
\left[ {\begin{array}{cc}
   \rho_{ee} & -\rho_{ge}/2 \\
   -\rho_{eg}/2 & -\rho_{ee} \\
  \end{array} } \right].
\label{rho_MOL_gamma_rad}
\eeq

The coupled Eqs. \ref{rho_CNT} and \ref{rho_MOL_gamma_rad} represent the quantum kinetic equation for the functionalized transport channel which we solve via the Runge-Kutta (RK4) method \cite{Runge} for time propagation.
We note that our derivation does not rely on perturbation theory expansion about the absorber-light field coupling. 

\renewcommand{\theequation}{B-\arabic{equation}} 
\setcounter{equation}{0}

\section*{Appendix B: Electron correlation effects within the $GW$ approxination}

In this section, we discuss the impact of electron correlation and image charge effects on our
results. Our approach for solving the quantum kinetic equation for the functionalized transport channel is based on the time-dependent Hartree-Fock (TDHF) level of theory. Since it is known that electron correlation effects beyond TDHF can be important \cite{Stefanucci,Myohanen}, we address below the question of whether these effects may alter the conclusion of our work, namely that single-photon detection can be achieved with a functionalized transport channel device under optimal operating conditions. Electron correlations effects play different roles in different parts of the functionalized transport channel:\\
\\
A) {\it Electron transport channel subspace}. The electronic properties of the transport channel are expected to be impacted by inclusion of electron correlation effects inside the channel subspace. These effects may alter the quasiparticle energy levels of the channel as well as their energy broadening, which may lead to significant changes in the channel current \cite{Stefanucci} especially when screening effects in the channel are important.\\
\\
B) {\it Interaction between channel and absorber}. Electron correlation effects between the transport channel and the absorber are also important as they are expected to affect the absorber optical level via image-charge effects.\\
\\
C) {\it Absorber subspace}. Electron correlation effects inside the absorber subspace are not important because no charge transfer is allowed between absorber and channel and the absorber only accommodates a single electron. As discussed in Appendix A, the absorber dynamics at the Hartree-Fock level is effectively equivalent to that of an isolated single-electron system subject to a modulated light field. Because in this case there are no Coulomb interactions inside the absorber subspace, treating the Coulomb interaction within first-order perturbation-theory ({\it i.e.} within Hartree-Fock) has no impact on the quantum dynamics of the coupled system -as numerically verified. Higher orders in perturbation theory inside this subspace are thus not expected to play an important role.

\subsection{Electron correlation effects in the transport channel}

We calculated the impact of electron correlations in the channel by including them at the $GW$ level \cite{Louie} during the steady-state calculations that determine the constant $c$ and function $f$ characterizing the absorber action/channel backaction effects discussed in the main text.
We compare the SNR results for 2 cases: 

- Neglecting Coulomb interactions in the channel ($U$ set to zero inside the channel). We have used this approximation throughout the results presented in the main text as it allows a gain in the computational time of several orders of magnitude over the case where Coulomb interactions in the channel are included. 

- Electron self-energy effects in the channel considered at the $GW$ level. We have implemented these effects by extending to the multi-site channel case a $GW$ approach \cite{Spataru_AM1,Spataru_AM2,Spataru_AM3} previously applied by one of us to the single-site channel case. In this case we considered a spin-dependent Coulomb interaction screened by an average screening constant \cite{Lanzara,Rotenberg} $\kappa=(\epsilon_{vacuum}+\epsilon_{substrate})/2$.  Assuming a conducting channel on an $Al_2O_3$ gate oxide we set $\epsilon_{substrate}=9$ which yields $\kappa=5$.

\begin{figure}
\includegraphics[trim=-50 0 -50 0,clip,width=\columnwidth]{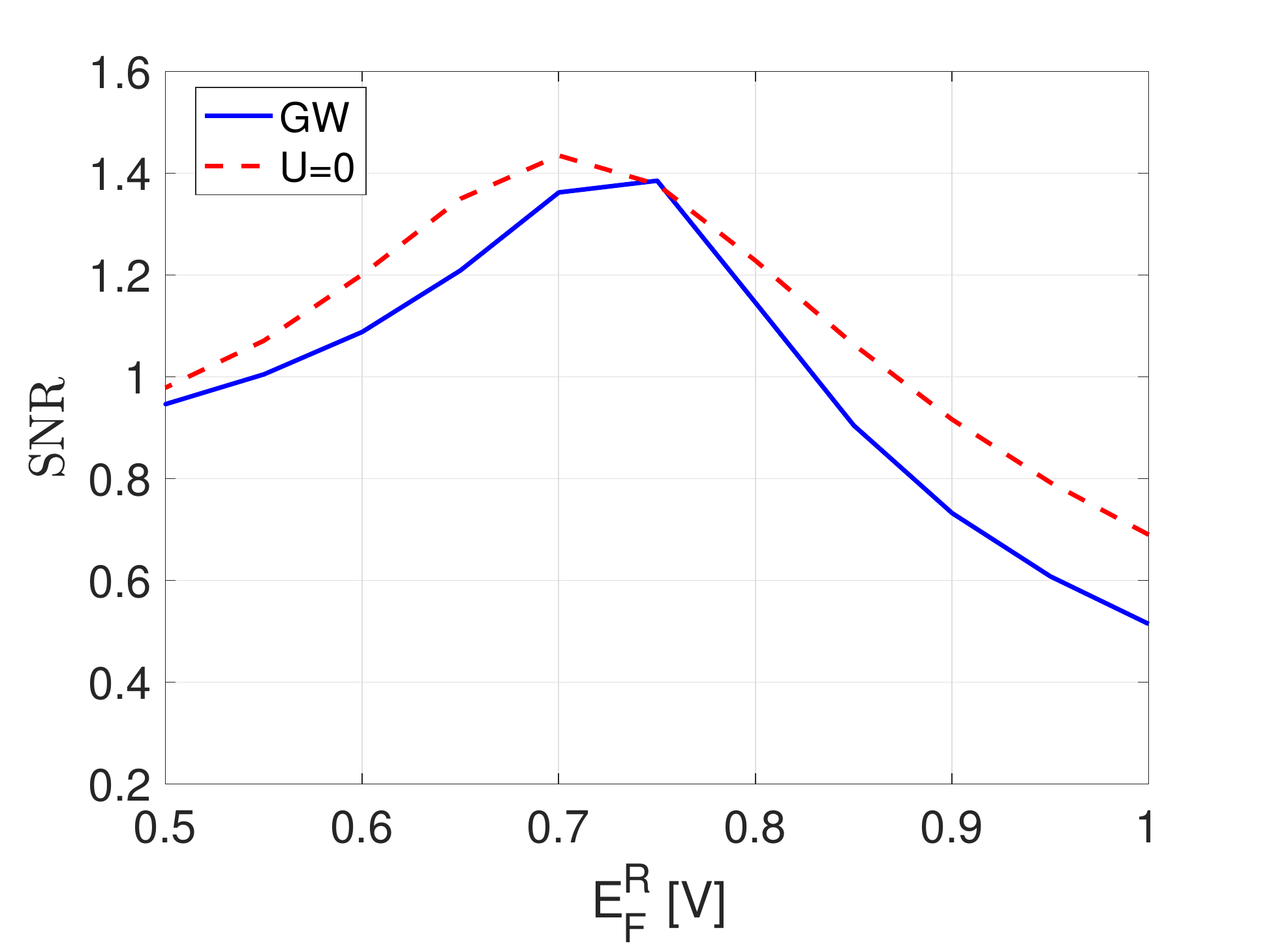}
\includegraphics[trim=-50 0 -50 0,clip,width=\columnwidth]{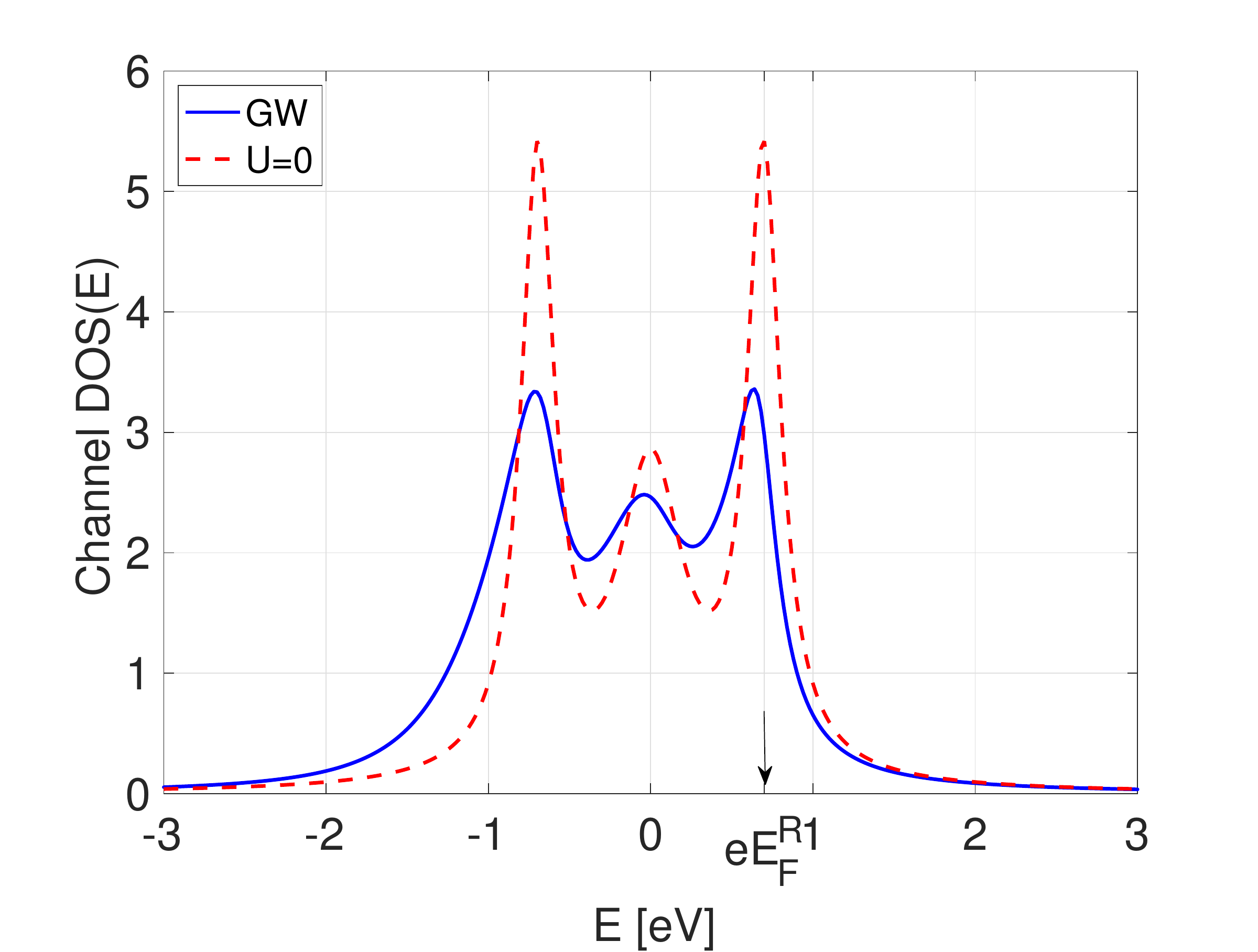}
\caption{(a) Optimized SNR for an absorber placed above the middle of a $3$-site channel. $E_F^L$ is well above the channel energy levels and the pulse duration $\tau_0$ is optimized for each $E_F^R$ value. The time window is $1.2$ ns and starts when the photon pulse begins interacting with the absorber. $\tau^{rad}=1$ ns. $N=3$, $j=2$, $d=20$ bohr. The Coulomb interaction inside the channel is either neglected ($U=0$) or treated at the $GW$ level. (b) Electronic DOS of the channel for the case of $E_F^R=0.7$ V (indicated by arrow).}
\vspace{-0.5cm}
\label{SNR_N3_GW}
\end{figure}

The SNR results for the two cases are shown in Fig.~\ref{SNR_N3_GW}(a) for an absorber above the central site of a 3-site conducting channel. The lead Fermi levels are chosen such that one of them ($E_F^L$) is well above the energy levels in the channel while the other one ($E_F^R$) is near the highest energy level: the optimal SNR is obtained precisely in this typical configuration where the doping level in the channel is small (one Fermi level pinning the highest peak in the electronic DOS of the channel -see Fig.~\ref{SNR_N3_GW}(b)).
The comparison shows that screening effects inside the conducting channel (that are captured at the $GW$ level) play a minor effect on the optimized SNR (with maximum value of about $1.4$), leading to an insignificant difference between the 2 cases. This is due to the relatively low doping levels (hence screening effects are small) as well as the fact that the induced changes (due to screening inside the channel) in the absorber action and channel backaction tend to cancel each other.
Based on these results we expect that the main conclusion of our work (that it is possible to achieve large SNRs under optimal conditions) is not affected by the neglect of Coulomb interactions inside the channel.

\subsection{Electron correlation effects in absorber-channel interactions}
\subsubsection{Image-charge effects in the linear-response regime}
In the linear-reponse regime ({\it {i.e.}} small absorber excitation probability \cite{lin-resp}) and for an absorber with permanent dipole in the excited state, the absorber optical level is renormalized via the image charge effect originating from the interaction between the absorber excited state {\it permanent} dipole and its image in the channel \cite{Thygesen,SpataruCNTmetal,SpataruCNTmetal2}. This effect 
is not captured within time-dependent Hartree-Fock and to account for it via steady-state calculations one needs to include electron correlations effects (e.g. within the $GW$ approximation) in the coupled absorber-channel space \cite{Neaton}. To good approximation it reads \cite{Thygesen}:
\beq
\Delta E_g^{im} \approx 1/2 \langle \phi_e^*\phi_e-\phi_g^*\phi_g|w_{ch}-v|\phi_e^*\phi_e-\phi_g^*\phi_g \rangle 
\label{GW_Kd}
\eeq
where $w_{ch}$ is the static screened Coulomb interaction of the channel alone and $w_{ch}-v$ represents the image potential due to the channel.

The image-charge effect described by $\Delta E_g^{im}$ affects the light-absorber resonant condition in the linear-response regime. However, capturing quantitatively the renormalized absorber optical gap in the linear-response regime is not essential since we assume in the main text that resonant light-absorber coupling is achieved by adjusting the photon energy to match the absorber optical gap. Adding the contributions $\Delta E_g^{im}$ would only mean that the
optimal SNR would occur at a slightly different photon energy, but would not affect the value of the SNR
or the dynamics of the system.
\subsubsection{Image-charge effects in the non-linear response regime}
The regime addressed by our simulations belongs to the non-linear response where additional {\it time-dependent} renormalization of the absorber optical gap occurs. It is critical to capture this dynamical effect as it determines the backaction from the channel to the absorber and affects the quantum dynamics of the system. There are two main contributions to the dynamical renormalization of the absorber optical gap:\\ \\
(I) The dynamical Stark effect, denoted in the main text as $\delta E_g(t)$;\\ \\
(II) Time-dependent contributions to $\Delta E_g^{im}$, which we denote here as $\delta E_g^{im}(t)$. 
\\ \\
We argue below that contribution (I) (captured by our simulations described in the main text) is the dominant one in the non-linear response regime.\\

For a qualitative comparison it is useful to express $\delta E_g(t)$ as an image charge effect, distinct from the one described in subsection B.1. To see this, consider the case where the absorber excitation probability has increased at time $t$ to  $\mathcal{P}(t)$. In practice $\mathcal{P}(t)$ is small (because of channel backaction it only reaches values of a few $\%$) hence in the analysis below it is sufficient to consider linear expansion about $\mathcal{P}(t)$. Since the absorber acquires a partial permanent dipole $\mathcal{P}(t)\times\mu_{perm}$ and an associated change in its charge distribution $\delta \rho_{abs}(t)=\mathcal{P}(t) [\phi_e^*\phi_e-\phi_g^*\phi_g]$, it creates an 'image' charge perturbation $\delta \rho_{ch}(t)$ in the conducting channel. 
This leads to a new type of absorber optical gap renormalization (w.r.t. $t=0$, {\it i.e.} w.r.t. the linear-response regime \cite{lin-resp2}) that originates from a change in the Hartree self-energy $\delta\Sigma^{Ha}_{abs}\equiv\Sigma^{Ha}_{abs}(\rho+\delta \rho_{ch})-\Sigma^{Ha}_{abs}(\rho)$.
To see this let's consider the renormalization of the HOMO/LUMO (g/e) absorber energy levels $\epsilon_{g/e}$:
\beq
\delta\epsilon_{g/e}(t)=\langle \phi_{g/e} | \delta\Sigma^{Ha}_{abs}(t) | \phi_{g/e} \rangle = \langle \delta\rho_{ch}(t) |v| \phi_{g,e}^*\phi_{g,e} \rangle.
\label{Delta_Hartree_QP}
\eeq
We use the fact that the interaction between a test charge distribution  $|\phi_{g,e}|^2$ and the charge distribution $\delta \rho_{abs}(t)$ in proximity to the transport channel is mediated by $w_{ch}$, leading to the electrostatic potential felt by the test charge distribution : $\langle \delta \rho_{abs}(t) | w_{ch}| |\phi_{g,e}|^2 \rangle$. Using the method of images this potential can be re-written as the sum between the electrostatic potential created by $\delta \rho_{abs}(t)$ alone $\langle \delta \rho_{abs}(t) | v | |\phi_{g,e}|^2 \rangle$ and the electrostatic potential created by the 'image' charge perturbation $\delta \rho_{ch}(t)$: $\langle \delta \rho_{ch}(t) | v | |\phi_{g,e}|^2 \rangle$. Equating these allows us to write
\begin{multline}
\delta\epsilon_{g/e}(t)=\langle \delta \rho_{abs}(t)) | w_{ch}-v | |\phi_{g,e}|^2 \rangle
=\\
\mathcal{P}(t)\langle |\phi_e|^2-|\phi_g|^2 |w_{ch}-v| |\phi_{g,e}|^2 \rangle
\nonumber
\end{multline}
and the corresponding renormalization of the absorber optical gap
\beq
\delta E_g(t) \equiv  \delta\epsilon_{e}(t) -\delta\epsilon_{g}(t) = \langle \delta\rho_{ch}(t) |v| |\phi_e|^2-|\phi_g|^2 \rangle
\label{Delta_Hartree_1}
\eeq
is written as
\beq
\delta E_g(t) = \mathcal{P}(t) \langle |\phi_e|^2-|\phi_g|^2 | w_{ch}-v||\phi_e|^2-|\phi_g|^2 \rangle.
\label{Delta_Hartree_2}
\eeq 

Let us focus next on the image-charge effect not included in our simulations $\delta E_g^{im}(t)$, and see how it compares with $\delta E_g(t)$. This change originates from the dynamical renormalization (due to $\delta \rho_{abs}(t)$) of the channel screened Coulomb interaction $\delta w_{ch}(t)\equiv w_{ch}(t)-w_{ch}$. Taking this into account Eq. \eqref{GW_Kd} yields
\beq
\delta E_g^{im}(t) \approx 1/2  \langle |\phi_e|^2-|\phi_g|^2|w_{ch}(t)-w_{ch}||\phi_e|^2-|\phi_g|^2 \rangle. 
\label{Delta_GW_Kd}
\eeq
We also expand $w_{ch}(t)$ to first order in $\mathcal{P}(t)$: $w_{ch}(t)=w_{ch}+\mathcal{P}(t)\dot{w}$ where $\dot{w}\equiv \frac{\delta w_{ch}(t)}{\delta \mathcal{P}(t)}$ is a generic quantity that depends on the absorber-channel separation $d$ and satisfies $\dot{w}(d\rightarrow\infty)=0$. This leads to:
\beq
\delta E_g^{im}(t) \approx 1/2 \mathcal{P}(t) \langle |\phi_e|^2-|\phi_g|^2|\dot{w}||\phi_e|^2-|\phi_g|^2 \rangle. 
\label{Delta_GW_Kd_2}
\eeq
By comparing Eqs \eqref{Delta_Hartree_2} and \eqref{Delta_GW_Kd_2} it becomes clear that the two terms display different scaling behavior w.r.t.~the absorber-channel separation $d$. Indeed, $w_{ch}$ on the r.h.s.~of Eq.~\eqref{Delta_Hartree_2} is obtained in the absence of the absorber hence it does not depend on $d$. By contrast $\dot{w}$ on the r.h.s.~of Eq.~\eqref{Delta_GW_Kd_2} becomes negligible at large $d$. \\

\begin{figure}
\includegraphics[trim=-40 0 -40 0,clip,width=\columnwidth]{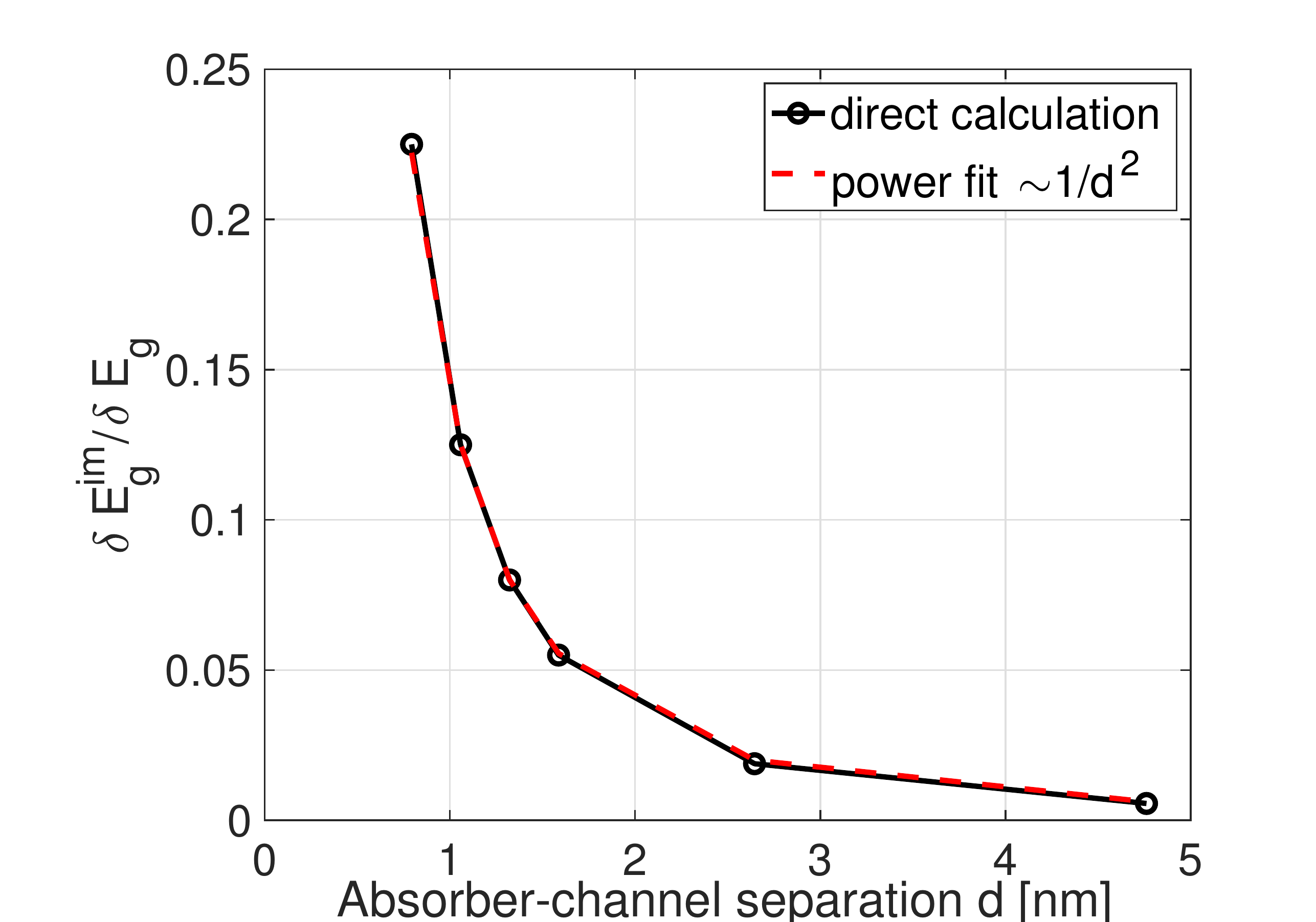}
\includegraphics[trim=-80 0 -50 0,clip,width=\columnwidth]{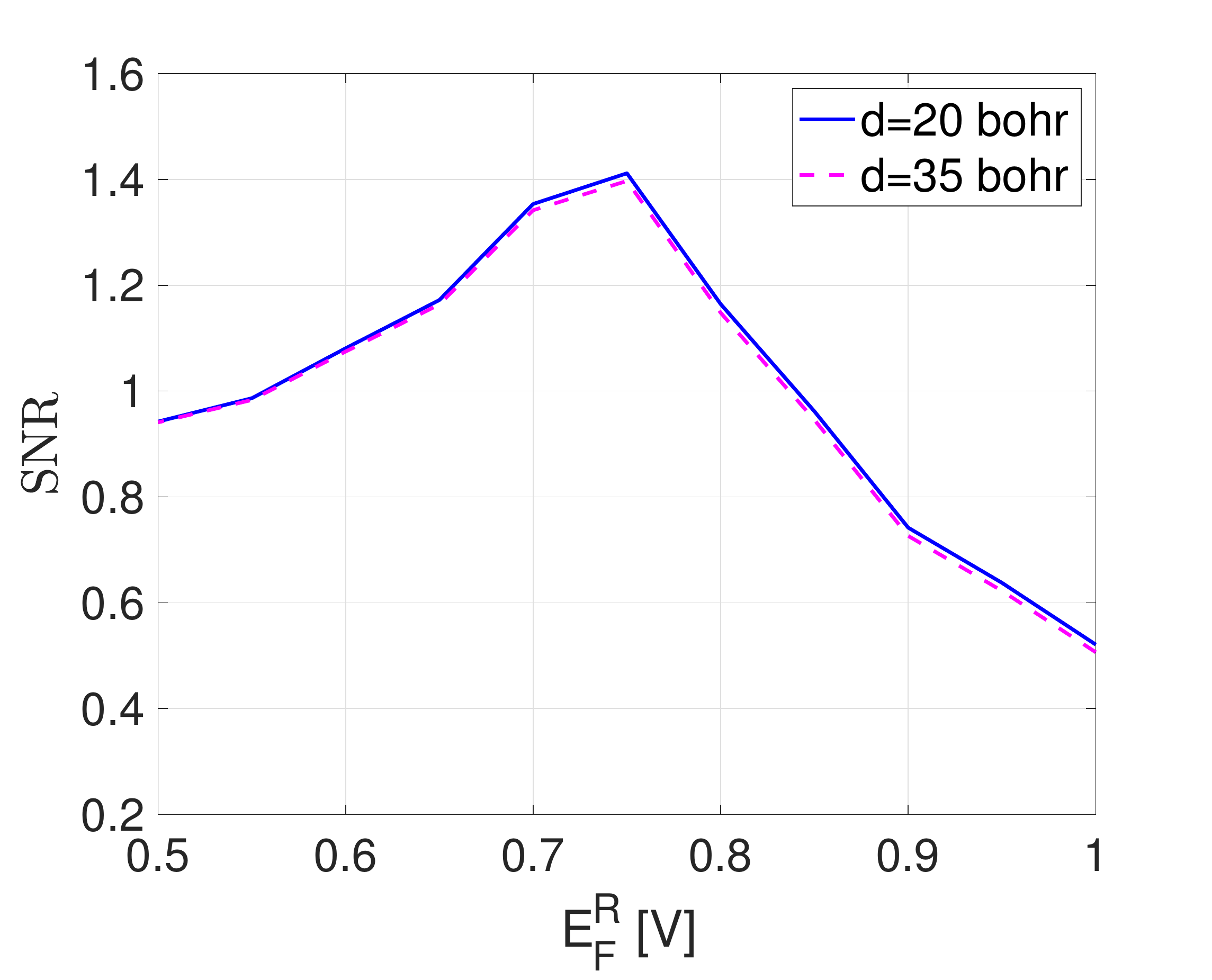}
\caption{(a) Ratio $\delta E_g^{im}(t)/ \delta E_g(t)$. $E_F^L$ is well above the channel energy levels and $E_F^R=0.7$ eV. (b) SNR results for two different absorber-channel separations: $d=20$ bohr (same as in fig. \ref{SNR_N3_GW}(a) - solid blue line) and $d=35$ bohr. $N=3$, $j=2$. $d$ is defined as the distance between the absorber site farthest away from the channel and the central channel site. Calculations are performed by treating the electron correlation inside the conducting channel at the $GW$ level.}
\vspace{-0.5cm}
\label{SNR_d_N3_ratio_dEg}
\end{figure}

For a quantitative comparison we calculated the two contributions $\delta E_g^{im}(t)$ and $\delta E_g(t)$ using Eqs \eqref{Delta_GW_Kd} and \eqref{Delta_Hartree_QP} for a 3-site channel with
source-drain and gate voltages that correspond to the optimal SNR for this geometry (see Fig.~\ref{SNR_N3_GW}). For consistency both terms were calculated with the electron correlation inside the conducting channel treated at the $GW$ level.
The calculated ratio $\delta E_g^{im}(t)/ \delta E_g(t)$ is independent of the particular (small) value of $\mathcal{P}(t)$ (as it should be from Eqs \eqref{Delta_Hartree_2} and \eqref{Delta_GW_Kd_2}) and the result is shown in Fig.~\ref{SNR_d_N3_ratio_dEg}(a). 
As expected  $\delta E_g^{im}(t)/ \delta E_g(t)\rightarrow 0$ at large $d$ (more exactly we find a $\sim1/d^2$ behavior).
We see that for the typical separation $d\sim1$ nm the term $\delta E_g^{im}(t)$ is about one order of magnitude smaller than the term $\delta E_g(t)$, {\it i.e.} it would lead to a $\sim10\%$ increase in the channel backaction. This $10\%$ increase in the backaction leads to less than $5\%$ decrease in the SNR \cite{action}. At larger separations $d> 1$ nm the impact of the term $\delta E_g^{im}(t)$ becomes even smaller. 

The above can be corroborated with the fact that the SNR (calculated without accounting for $\delta E_g^{im}(t)$) is virtually independent of $d$, as seen in Fig.~\ref{SNR_d_N3_ratio_dEg}(b). This is due to the fact that with increasing absorber-channel separation the reduced Coulomb coupling leads to a decrease in channel backaction that is almost exactly canceled by a corresponding decrease in the absorber action. We note that with increasing distance one adjusts accordingly the pulse duration: the optimal $\tau_0$ increases with $d$, reflecting the smaller channel backaction. The robustness of the SNR against increasing distance holds until the optimal $\tau_0$ saturates - when reaching values $\sim\tau_{rad}$.

Together, the results in Fig.~\ref{SNR_d_N3_ratio_dEg} lead us
to conclude that the term $\delta E_g(t)$ (accounted for in the results presented in the main text) is the dominant renormalization term for $d\gtrsim 1$ nm and  that the main conclusion of our work (that it is possible to achieve large SNRs under optimal conditions) is not affected by the neglect of contribution (II).

\begin{acknowledgments}
We thank Mohan Sarovar for useful discussions. Work supported by the Defense Advanced Research Projects Agency (DARPA) DETECT program. The views, opinions and/or findings expressed are those of the author and should not be interpreted as representing the official views or policies of the Department of Defense or the U.S. Government. Sandia National Laboratories is a multimission laboratory managed and operated by National Technology and Engineering Solutions of Sandia, LLC., a wholly owned subsidiary of Honeywell International, Inc., for the U.S. Department of Energy's National Nuclear Security Administration under contract DE-NA-0003525. 
\end{acknowledgments}



\begin{thebibliography}{99}
\bibitem{substrate_loss} Additional care needs to be paid when designing a realistic device where the conducting channel is placed on top of a substrate: the electronic level alignment and/or separation between absorber and substrate need to be judiciously chosen to minimize energy transfer losses.
\bibitem{other_approaches} {For the particular case of a coherent light pulse the OBE approach is equivalent 
to more general approaches such as the quantized propagating pulse \cite{Scarani} or master equation \cite{Combes} approaches.}
\bibitem{Fano}  For $N=7$ and the design that yields SNR $\approx 4.5$ we estimate $F=0.35$ from 
$F=\int dE (f_L-f_R)T(1-T)$ / $\int dE (f_L-f_R)T$ where $T(E)$ is the calculated Landauer transmission. This estimate however does not account for the impact of the absorber on current fluctuations.
\bibitem{HF} 
In practice, the NEGF dynamics is obtained by treating the interaction between the absorber and the transport channel at the Hartree level (the exchange self-energy vanishes in the absence of charge-transfer between absorber and transport channel). Because the absorber accommodates a single electron there are no Coulomb interaction effects that need to be accounted for inside the absorber subspace. Inside the transport channel subspace we include spin-degeneracy but neglect Coulomb interaction effects. The impact of electron correlation effects beyond the Hartree-Fock level is studied within the $GW$ approximation in Appendix B.
\bibitem{lin-resp} By that we mean that the excited dipole is small enough such that the renormalization of the absorber optical gap remains well within the bounds set by the photon bandwidth (see inset of Fig. 2 in the main text).
\bibitem{lin-resp2} By $t=0$ we mean the initial times (when the photon field starts interacting with the absorber) where the renormalization of the absorber optical gap can be neglected. The corresponding timescale is of course larger than the $\sim$fs light period.
\end{thebibliography}
\end{document}